# ECG Encryption and Identification based Security Solution on the Zynq SoC for Connected Health Systems

Xiaojun Zhai, Department of Electronics, Computing and Mathematics, University of Derby, UK

Amine Ait Si Ali, KINDI Center for Computing Research, Qatar University, Qatar

Abbes Amira, KINDI Center for Computing Research, Qatar University, Qatar

Faycal Bensaali, KINDI Center for Computing Research, Qatar University, Qatar

Corresponding author: Xiaojun Zhai, Department of Electronics, Computing and Mathematics, Kedleston Road, Derby, DE22 1GB, Tel: 01332593892, Email: x.zhai@derby.ac.uk

*Abstract*— Connected health is a technology that associates medical devices, security devices and communication technologies. It enables patients to be monitored and treated remotely from their home. Patients' data and medical records within a connected health system should be securely stored and transmitted for further analysis and diagnosis. This paper presents a set of security solutions that can be deployed in a connected health environment, which includes the advanced encryption standard (AES) algorithm and electrocardiogram (ECG) identification system. Efficient System-on-Chip (SoC) implementations for the proposed algorithms have been carried out on the Xilinx ZC702 prototyping board. The Achieved hardware implementation results have shown that the proposed AES and ECG identification based system met the real-time requirements and outperformed existing field programmable gate array (FPGA)-based systems in different key performance metrics such as processing time, hardware resources and power consumption. The proposed systems can process an ECG sample in 10.71 ms and uses only 30% of the available hardware resources with a power consumption of 107 mW.

## I.  INTRODUCTION

Population of middle aged and old people is the most dominated in the highest developed countries and regions, which requires governments to deal with the problems in the health-care sector [1]. This results on limited number of working adults to take care of the growing dependent elderly population, which may cause potential financial problems, but also increasing the time for a patient to receive treatment [2]. Therefore, new solutions are necessary to increase the level of automation from the existing systems and be able to safely and efficiently handle the enormous amounts of data generated, stored and transmitted between them [3]. One of the most promising technologies to achieve this is to apply the Internet of Things (IoT) paradigm, in which information and communication systems are embedded in a health care environment [4]. In this paper, the field of wireless monitoring of vital signs using IoT is proposed, in which it embraces the measuring and digitisation of vital signs such as the blood pressure or electrocardiograms (ECGs), transmitting packets over a wireless network and delivering this medical information to health-care professionals. It allows the use of what is defined as pervasive health-care: "health-care to anyone, anytime and anywhere" [5]. It can be used in clinics and hospitals to ease the monitoring of patients, but also outside of medical facilities, giving elderly or sick people the opportunity to be part of their social communities by being simultaneously monitored and/or even in contact with their health-care provider. Such systems can, if they are designed appropriately, deal with a high amount of patients, by consuming fewer resources in terms of care attendants and capacities of medical facilities and help to make health-care more efficient and economical.

Since the health-care data contains highly sensitive and personal data, strong security issues should be addressed to avoid serious consequences, causing damage, disruption to operations or, in some scenarios, even loss of life, one of the solution to that is to apply protection and encryption on the healthcare data [6]. There is a high interest in using the health-care monitored biometric signals to identify the individual patients. The main benefits of this approach are that there is no need for supplying extra biometric sensors and the biometric signals are continuously monitored together with other medical signals. An ECG signal is less prone to fraud and it has been widely used in the field of health monitoring where it can provide an automatic living identification system. This system helps monitoring the life of the patient since the ECG signals are a sign that the person is still alive and provide vital information about the health conditions. The drawback of using ECG signals for human identification is that there is a large number of signals coming from the sensor that need to be processed in real-time.

In addition to patient identification, privacy and security of transmitting and storing patient's medical data is also a major issue in connected health. Data and various information are shared between different organizations, platforms and people in the health industry. Therefore, data encryption is vital to protect and secure information. Advanced encryption standard (AES) algorithm is considered as the state of the art encryption system that is widely used in many applications including connected health systems [7].

Programmable system-on-Chip (SoC) based implementations are being used by researchers to accelerate digital signal processing (DSP) algorithms to meet the real-time requirements by exploiting the parallelism, pipelining and the hardware/software co-design offered by such platforms. For the encryption and ECG identification systems, the SoC-based solution can be used to collect data from ECG sensors and process it in real-time. Compactness and cost effectiveness are some of the advantages when using such solutions. Furthermore, reconfigurability of programmable devices provides the user with a possibility to easily upgrade and calibrate their systems depending on the needs. Heterogeneous platforms, such as the Zynq SoC, based hardware, provide not only similar advantages but they also provide the user with higher flexibility where various interfaces, processing system (PS) and programmable logic (PL) can be used.

This paper presents a set of security solutions that can be deployed in a connected health environment with application to ECG. This includes ECG encryption using the AES algorithm and ECG identification using multiresolution and principle component analysis (PCA). Efficient SoC implementations for the proposed algorithms have been carried out on the Xilinx ZC702 prototyping board equipped with the Zynq SoC device. The implementation is verified using the data obtained from two personal data sets measured from the VS100 ECG sensor [8] and the Shimmer3 ECG sensor [9] as well as the public ECG MIT-BIH database [10]. Achieved results have shown that the proposed system only requires 30% of the available hardware resources and 107 mW to process an ECG sample in 10.71 ms, which outperforms the existing field programmable gate array (FPGA)-based implementations in different key performance metrics. This paper starts by reviewing the state of the art AES algorithms and ECG identification systems in section II. The simulation of the proposed approach for AES algorithms and ECG identification are presented subsequently in section III. This is followed by the description of the proposed system architecture and its hardware implementation in Section IV. Experimental setup and results analysis are then given in section V. Finally conclusions are drawn.

## II. RELATED WORK

Connected health systems have been increasingly attracting many researchers. In this section, the most recent and related work to connected health is summarised. In [11], a system is proposed to monitor the heart of cardiac patients using ECG measurements. The heart electrical impulses are collected using an ECG sensor and a personal digital assistant (PDA) is used to process the ECG signals and perform the diagnosis. The PDA is capable of calling the medical staff in the case of an emergency or critical situation. The user can also visualize his/her own medical data using a graphical interface on the PDA. Another connected health system is presented in [12] where a sensorized glove for measuring hand finger flexion for rehabilitation purposes is developed. The system consists of a glove and a set of configured sensors placed in specific places and connected to the acquisition unit. The aim is to give feedback to the rehabilitation system using fingers' positions. In both [11] and [12] the security and privacy is to be improved.

AES/Rijndael is a block cipher that can encrypts blocks with a fixed length of 128 bits and a corresponding decipher that is basically a simple inversion of the cipher [13]. A high throughput pipelined approach was presented in [14], however, this was achieved with the sacrifice of high design effort and resource consumption. A similar low area and memory free solution for the implementation of the AES algorithm on FPGA was presented in [15], the proposed design that uses an 8-bit data path, supports 128-bit keys and requires 160 clock cycles for one encryption was implemented on both Spartan 3 (XC3S50) and Spartan 6 (XC6SLX4) FPGAs requiring 184 and 80 slices respectively. The claimed throughput is 36.5 Mbps for the Spartan 3 while it reaches 58.13 Mbps for the Spartan 6. The implemented design consists of five blocks: ShiftRow, Sbox, MixColumn, KeySchedule and Delay, the key point of this solution is that it uses the Xilinx SRL 16/32 to implement keySchedule block to reduce the number of slices. Another high throughput and pipelined implementation of AES algorithm on FPGA was presented in [16], the implementation was performed on both Virtex-5 and Virtex-6 and the best claimed throughput of 260 Gbit/s was achieved for the implementation using the counter mode on the Virtex-6. The key point is that the Sbox implementation here combines between both memory and non-memory based approaches. A high throughput solution was also presented in [17] for an implementation on Virtex 5 using VHDL, the claimed throughput is 222 Gbit/s. Other solution was presented in [18] for the implementation of the AES encryption algorithm on an ASIC 65-nm CMOS, 22-nm CMOS and NVidia GeForce 8800 GTX. A hardware implementation on FPGA of an improved version of the unsecure data encryption standard (DES) is presented in [19]. It has been presented in the context of threats to cryptographic chips. The literature review has also shown that the use of high level synthesis (HLS) tools such as Vivado for the implementation of the AES algorithm in contrast with the use of HDL codes can help resolving various issues such as complexity and the time required for design verification and evaluation, this has been discussed in [20] and [21]. It has been shown that HDL based design offers more flexibility and control while using HLS based design save a considerable amount of time while having the same frequency, throughput and area.

Various techniques have been used in the past along with different implementations for the use of ECG signals for human identification. Two main approaches are being used to extract the most useful features from signals generated by an ECG sensor [22]. The first one is the fiducial approach that consists in the search for specific points of interest to extrapolate timing and magnitude measurements. In this approach the typical features are linked to peaks and timing duration of the P, QRS and T waves which are the representation of the heart activity in terms of depolarization and repolarization of the atria and ventricles. Features related to the physical functionality of the heart are not considered in the fiducial approach, only statistical and analytical features from the morphology of the signal waveforms are considered. In [23], a fiducial approach based on the multi resolution Daubechies D4 and D6 wavelet transforms is proposed to detect the QRS complex and the onsets and offsets of the P and T waves on the MIT-BIH database [10]. A positive recognition rate of 98% was achieved. In [24], researchers have developed a quadratic Spline wavelet based framework that aims at the automatic analysis of single lead ECGs for human identification, the maxima, minima and zero crossing values in the wavelet coefficient reconstruction of the ECG signals at various scales are found prior to the detection of the QRS, P and T fiducial points. The fiducial approach in this contribution has shown a positive recognition of 99.61% when applied to the MIT-BIH database. The second main approach to extract features from ECG signals for human identification is called the fiducial independent approach. The fiducial independent approach was used by [25] achieving a positive recognition rate of 99.6%. The wavelet coefficients have been extracted by using the Daubechies wavelet of order eight, the independent component analysis (ICA) is used to find the independent components from the statistical independent random variables; finally the PCA is used to reduce the dimensionality of the extracted feature. Various SoC based human identification systems using ECG signals have been found in the literature [26-29]. An ECG-QRS complex detection has been implemented on the Xilinx Virtex-II pro FPGA in [26], an identification rate of 99.681 % was achieved testing on the MIT-BIH database. Another SoC implementation using FPGA was presented in [27], it consists in the detection of arrhythmia patterns using ECG signals. In [28] and [29], two other embedded system based implementations were presented for the ECG signals denoising, filtering and compression.

The work presented in this paper aims at providing an efficient architecture and implementation of the AES-128 encryption and ECG-based identification system for protecting personal medical data privacy. The achieved results show that the speed, resources and power consumption of the proposed implementation are sufficient for the use of the proposed architectures in the connected health system. The main contributions of this paper can be summarised as follows:

A novel SoC solution is introduced. It unifies wireless health-care monitoring system with the identification of individuals using ECG.

The proposed approach introduces a way to integrate the acquisition and processing unit into reconfigurable hardware. This allows the implementation of a high-performance state-of-the-art data processing system which is also highly adaptive. The communication, visualization, security and identification can be realized on one piece of hardware without making the compromise of resource sharing and time-consuming sequential execution of tasks.

## III. PROPOSED SYSTEM

### A. System Overview

An overview of the proposed system can be seen in Fig. 1. The aim is to collect ECG signals in a safe and secure environment such as a hospital, home or ambulance. The collected data can then be sent wirelessly to the local processing unit based on the Zynq SoC where all information will be processed in real-time. Identification of the patient is performed based on the ECG signals collected locally, other ECG based examinations can be performed in the hospital or in the ambulance In the case where the data need to be sent to a different location for further examination or for exchange of information, the data will be encrypted first using the AES algorithm and then decrypted when received in the final destination. Three types of ECG signals data sets have been used to evaluate and validate the system. Two are private data sets collected in the lab and one is a public data set. The two personal data sets are obtained using the VS100 ECG sensor [8] and the Shimmer3 ECG sensor [9] while the public one is the MIT-BIH database [10].

### B. Software Simulation

Simulation of the proposed algorithms is carried out using both MATLAB and C/C++ environment, where MATLAB is used to generate an ECG database that contains the required files for testing the AES cipher-decipher block and ECG identification block. Fig. 2 shows an overall system block diagram.

*1) ECG Database*

In Fig. 2, the ECG database block consists of four different data, including original ECG signal, Eigen ECG vector, projected training vector and mean vector. The original ECG signal is divided into two sub-sets, one set is used for training, and the other set is used for testing. The testing set is stored into text files, to be used in cipher block. However, the training file is used for enrolment and generating the relevant feature data sets. The Eigen ECG matrix $E$, projected training $p$ and mean vectors $m$ are the corresponding ECG features. It first establishes one single mean signal and then it calculates the variation of each ECG from the mean ECG signal and stores them again in a matrix $A$. This matrix is used to calculate the covariance matrix to reduce the computation time and resources. A surrogate $L$ of the covariance matrix $C$ is used to calculate the eigenvectors $V$ and eigenvalues $D$. The advantage of doing this is that $L$ has much lower dimension than $C$, which could significantly reduce the computation time. The eigenvectors are sorted and eliminated by the size of the eigenvalues. If the eigenvalue is lower than one, the corresponding eigenvector is eliminated. However, these eigenvalues do not represent the eigenvectors of the covariance matrix $C$, which are recovered by multiplying the sorted eigenvectors of $L$ with the deviation matrix $A$.

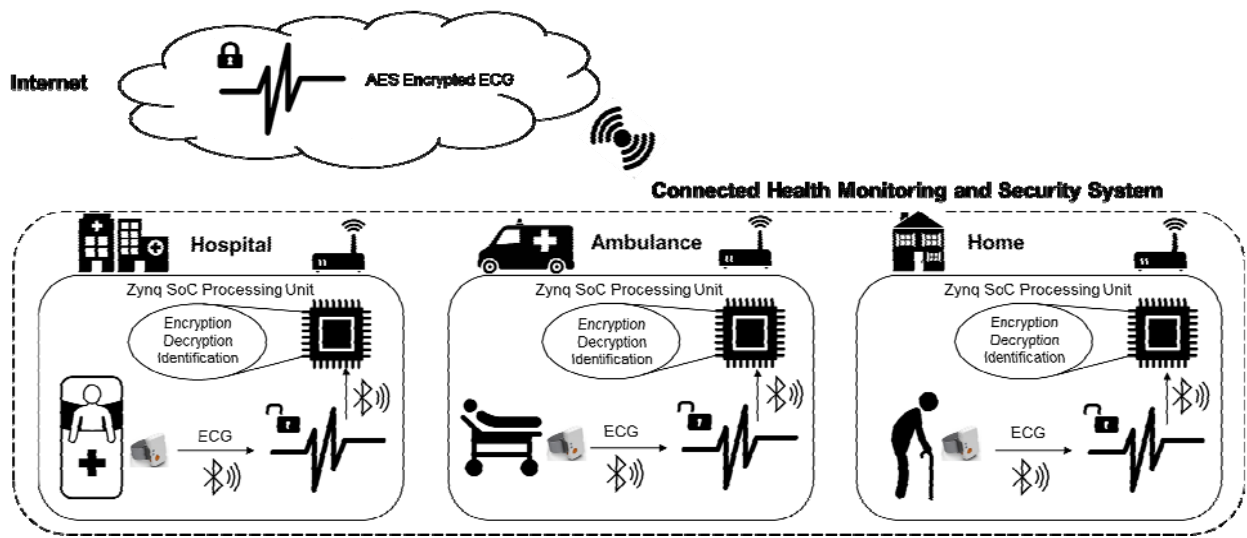

Fig. 1. System Overview.

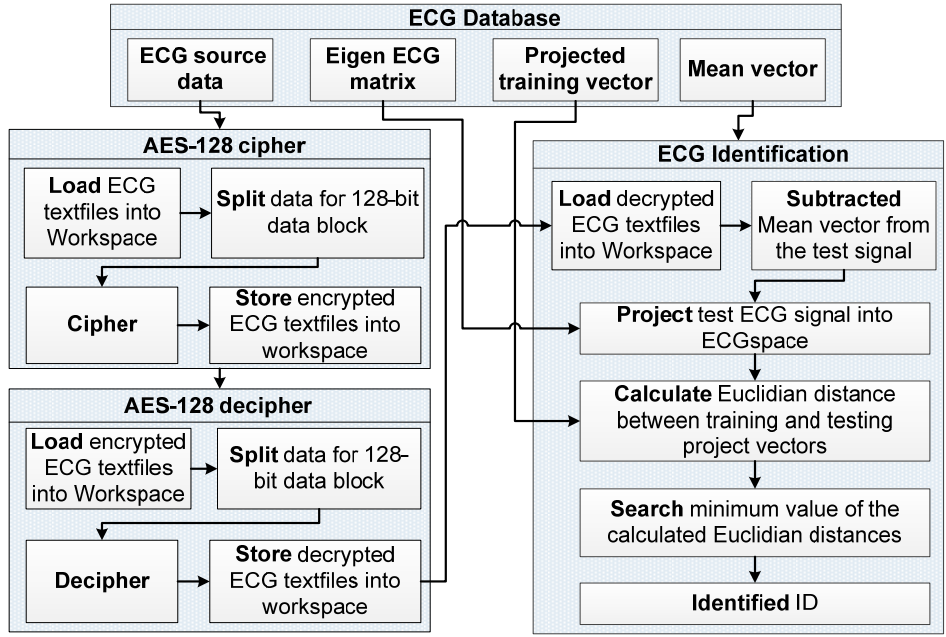

Fig. 2. Overall block diagram [30].

The deviated ECG signals are projected from matrix *A* onto the ECG space, which represents the unique features of the training ECG signals. In order to calculate the features of the ECG signals used for testing, the mean of the training database must first be subtracted from the test signal. Subsequently, the projected vector is then calculated via a multiplication of the Eigen ECG matrix by the mean subtracted signal. This process is represented by the following equation:

$$p = E^T \times I \qquad (1)$$

where *E* is the Eigen ECG matrix, *I* is the mean subtracted signal, and *p* is projected test vector.

2) *AES Cipher-decipher*

The text files that contain the original ECG signals are firstly used in the cipher block. The encrypted text files are then used in decipher block. Finally, the decrypted text files are used in the identification block.

The algorithms of the cipher-decipher block is revised and implemented in C++ based on the work presented in [31]. The algorithm supports AES-128 standard,

3) *ECG Identification*

In the ECG identification block, the same process of generating the database is repeated, where the testing ECG signal is projected to the ECG space using equation (1).

The next step is to calculate the Euclidian distance between each projected training vector and the projected test vector using equation 2. The index of the minimum error between the feature vectors represents the ID of the identified ECG.

$$d_i = \sqrt{\sum_{i=1}^{n}(p - p'_i)^2} \qquad (2)$$

where $d_i$ is the Euclidian distance between the $i^{th}$ training vector and the projected test vector. $p$ and $p'$ are the projected test vector and the $i^{th}$ projected training vector respectively.

This process could be called just once for a single identification or for all test samples for evaluating the algorithm. The single identification process is justified as it allows greater scrutiny of a test signal that was falsely recognized, because a plot shows three signals: the test ECG; the expected training ECG; and the identified ECG for further analysis, such as the difference between the signals. Furthermore, the single identification process can be used for security identification, as the displayed ID can be used for verification.

## IV. Hardware Architecture and Implementation

In order to accelerate the computational intensive part of the proposed algorithms a hardware architecture has been designed. The proposed architecture consists of three parts: cipher, decipher and ECG identification blocks. Vivado HLS [32] is used to design the three blocks, where C/C++ codes of the blocks are synthesised and translated to a hardware description language (HDL). In order to transfer data and control between each block, the interface of each block is designed to use AXI-lite Slave Peripheral interface, which means that each block can be accessed by a microprocessor through its peripheral bus. The main idea of this work is to design an architecture that could select different hardware acceleration blocks and perform computationally intensive algorithmic calculation, e.g. cipher-decipher and ECG identification block. A block diagram of the proposed system is shown in Fig. 3.

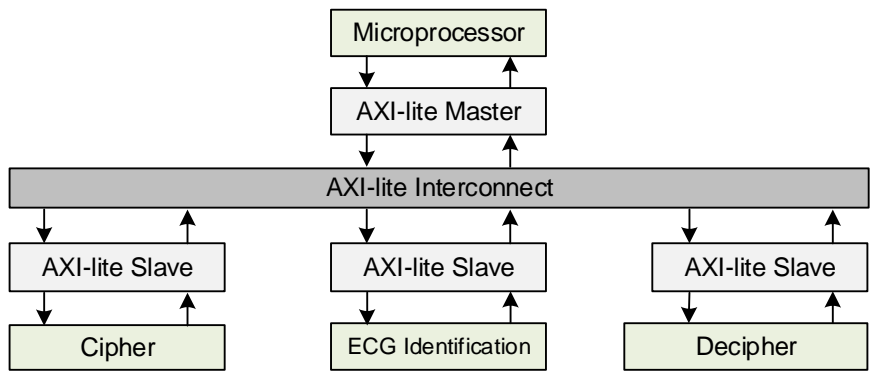

Fig. 3. A block diagram of the proposed system on the Zynq SoC.

## A. Cipher and Decipher Blocks

There are two data interfaces in the cipher block, which include an input and output data array. Both input and output data arrays are designed as 16×8-bit arrays, thus the block size for each calculation is 128 bits. The decipher block is an inverse calculation of the cipher block, which has the same input and output arrays as the cipher block. The algorithm of the cipher and decipher blocks have been implemented using C++. Directives in Vivado HLS to optimise the codes for hardware implementation have been used. The used directives are summarised in the following Table I.

TABLE I

USED DIRECTIVES IN CIPHER BLOCK

| **Instance** | Directives | **Function** |
|---|---|---|
| Interface | s_axilite | Create AXI-Lite slave interface |
| Array | Array_Reshape | Reshape the targeted array to a specific dimension |
| Loop | unroll | Transforms loop by creating multiple copies of the loop body |

Basically, by applying different directives from Table I, the architecture is synthesised under the user control. For example, the "set_directive_loop_unroll" command allows the loop to be fully unrolled or partially unrolled by a factor, creating as many copies of the loop-body in the register transfer level (RTL) as there are loop iterations. As a result of this, different iterations of the original loop can be run at the same time, thus the processing time of the block is significantly reduced.

## B. ECG Identification Block

Fig. 4 shows an overall diagram of the proposed ECG identification architecture. The proposed ECG identification architecture consists of two parts: the enrolment and hardware acceleration part. The enrolment part is completed on software; however, the hardware acceleration part is performed on the Zynq PL, which mainly consists of PCA values and Euclidian Distance calculators. The PCA values calculator is mainly used to perform the PCA projection (i.e. equation 1). The Euclidean Distance calculator is used to calculate Euclidean distance (i.e. equation 2).

### 1) PCA Projection Block

In the PCA projection block, the test input signal is first subtracted from the mean signal of the complete training matrix **T**. Subsequently, the projected vector is then calculated via a multiplication of the Eigen ECG matrix by the mean subtracted signal. This process is shown in Fig. 5.

In Fig. 5, the size of test and mean signal vectors is $n$, which is the length of the training vector. In this work, the length of the training vector is set to 300. The size of Eigen ECG matrix is $m \times n$, where $m$ is the number of PCA features, and $n$ is the size of test vector. The test signal is subtracted from the mean signal, and then the resulting vector is multiplied by each row of the Eigen ECG matrix one by one. After that, the results are accumulated and formed a project PCA vector with size of $m$.

*2) Euclidean Distance Calculator Block*

In the Euclidian distance calculator block, Euclidian distance of the projected training vector and the projected test vector is calculated as stated in equation 2. Since the calculation of the square root would not affect the search of the minimum value of the Euclidian distance, this calculation has been eliminated to reduce the hardware usage. Fig. 6 demonstrates the architecture of the Euclidian distance calculator block.

In Fig. 6, the projected test vector is the output of the PCA projection block, which has a size of m. The size of the projected training matrix is $m \times i$, where $i$ is the number of the training vectors in the training database. Each row of the projected training matrix is subtracted from the projected test vector, and then the resulting difference is multiplied by itself. Subsequently, an accumulator is used to sum all elements and form a Euclidian distance vector with size of $i$.

Once the calculated Euclidean distance vector is available, the next step is to search for the minimal Euclidian distance that represents the ID of the identified ECG.

*3) Directives Used in ECG Identification Block*

Similar to other two blocks, the interface directives used in the ECG identification block is still "AXI-lite slave bus". There are four input arrays, which are corresponding to the four parameters in the database block, shown in Fig. 2, respectively. However, the output of this block is only a variable that contains the identified ID. Unlike the used optimization in most loops used in the other two blocks, the directives "pipelined" is used in the ECG identification block, which could increase the throughput of the PCA projection and Euclidian distance calculator blocks.

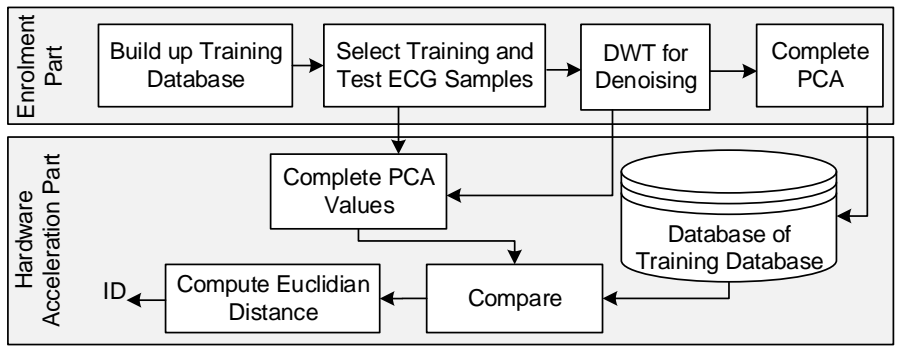

Fig. 4. An overall diagram of the proposed ECG identification system.

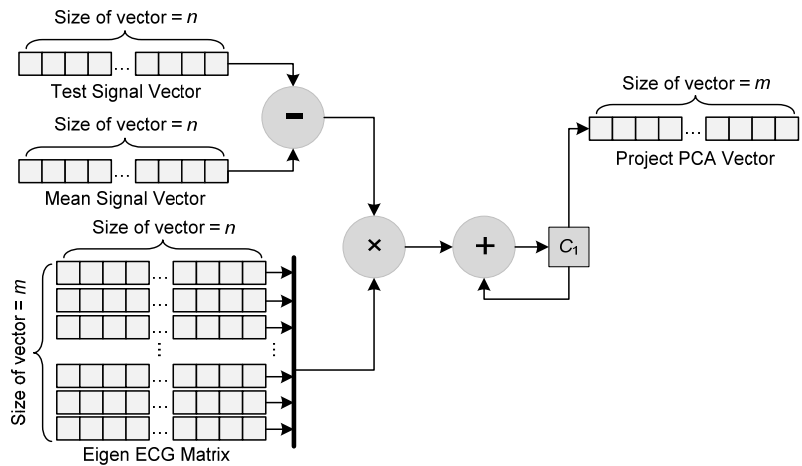

Fig. 5. Block diagram of the PCA projection block.

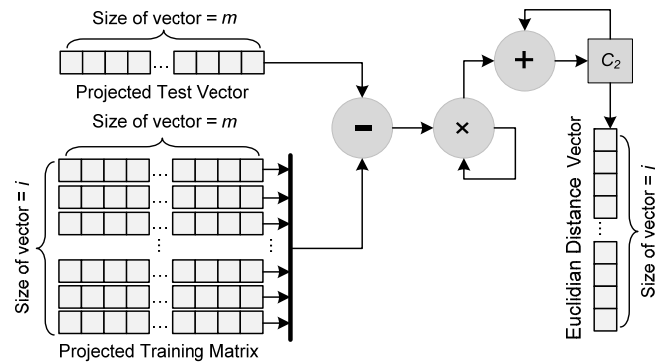

Fig. 6. Block diagram of Euclidean Distance Calculator.

*C. Implementation of the Proposed Architectures*

Xilinx Zynq programmable SoC is used for the implementation of the proposed architectures. Since this type of platform tightly integrates dual-core ARM Cortex-A9 processors with Xilinx 7-series FPGA logic [33], which makes it a perfect implementation platform of the proposed architecture. The proposed architectures were firstly generated as a set of individual IP catalogs using Vivado HLS, they were then integrated together using Vivado Design Suite [32]. Fig. 7 shows a block diagram of the proposed system with all building blocks and their interconnections generated by Vivado Design Suite as well as the chip layout highlighting the amount of resources used by each block.

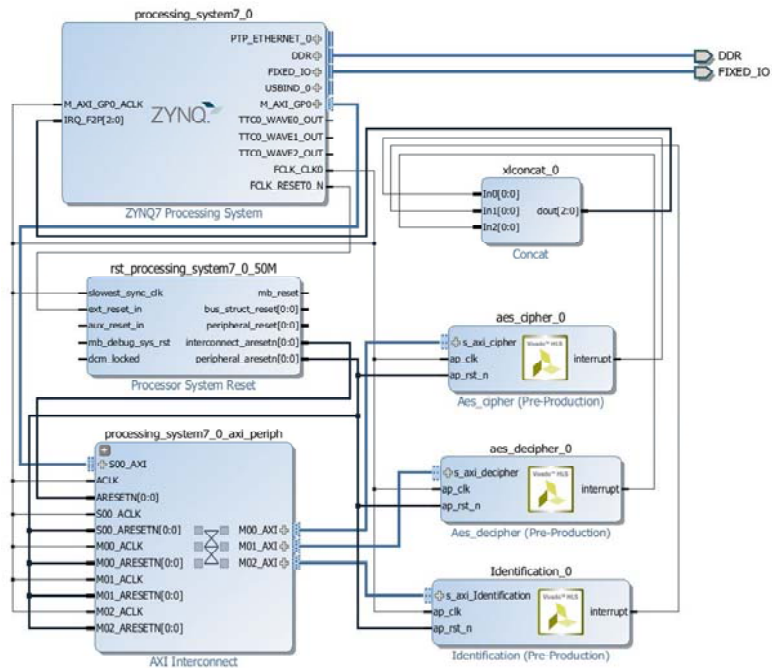

Fig. 7. Block diagram of the proposed implementation in Vivado Design Suite.

In Fig. 7, in addition to the three user IP Catalogs, there are four other Xilinx build-in IP Catalogs. They are "Xilinx Zynq7 Processing System", "Processor System Reset", "AXI Interconnect" and "Concat". The Xilinx processing system has direct physical *connections* to DDR memory, which is used for the running of the software in the Zynq7 processing system. A set of blocks of custom logic in the PL can be controlled and monitored by using memory mapped registers which can be accessed by processors via the AXI4 interconnect. The "Processor System Reset" is used to support asynchronous external reset input which is synchronized with the clock. The "Concat" IP is used to concatenate the interrupt signals generated from the different blocks of custom logic.

The software runs on the Zynq7 processing system that controls and drives the custom logic through AXI4 bus interface. The control process is summarized in the "Controlling the custom logic" pseudo code.

---

**Controlling the custom logic:**

1. Input: $D_{in}$ is the input data array of the the custom logic block.
2. Output: $D_{out}$ is the output data array of the custom logic block.
3. Initial processor and the custom logic;
4. **if** Initialization is successful **then**
5.    setup the interrupt;
6.   **if** setup the interrupt is successful **then**
7.     **for all** *data* in $D_{in}$ **do**
8.       write data into the custom logic;

| | |
|---|---|
| 9. | **if** the custom logic is ready **then** |
| 10. | start the custom logic; |
| 11. | **while** result of the custom is **not** ready **then** None**; end** |
| 12. | read data from the custom logic to $D_{out}$. |
| 13. | **end** |
| 14. | **end** |
| 15. | **end** |
| 16. **end** | |

## V. Experimental Setup and Result Analysis

In order to achieve real-time performance, the signal processing algorithms are implemented on PL as an ad-hoc digital circuit, which could be one of the valuable solutions for accelerating computationally intensive algorithms. In addition, the PL could also balance the gap between software and hardware design to allow maximum performance and flexibility to be delivered during development.

Xilinx Vivado HLS tool [32] has been used for the design and development of the proposed hardware architecture. The design was first implemented using C++, and then a C++ level simulation was performed. The purpose of this is to evaluate the results of the algorithm which should be the same results obtained from MATLAB implementation. After that, a C/C++ synthesis was performed to translate the codes to a HDL. VHDL was selected as the target HDL. Thereafter, a RTL simulation was employed where the same C++ testbench used in C++ level simulation has been used again to evaluate the final RTL implementation, which simplifies the design process for evaluating the signal processing algorithm.

### A. Vivado HLS Simulation

#### 1) C/C++ Simulation

Prior to the hardware implementation, the proposed cipher-decipher and ECG identification system was validated using Vivado HLS C simulator. ECG signals obtained from VS100 ECG sensor [8], Shimmer3 ECG sensor [9] and MIT-BIH database [10] were used for the evaluation. A total of 60, 2261 and 20 ECG signal samples from the VS, Shimmer3 and MIT-BIH databases have been used for testing respectively. The archieved recognition rate for each database is 99.5%, 95% and 100% respectively. Fig. 8 shows some samples from the used ECG databases. Once the codes passed Vivado C simulation, the C++ codes were translated to HDL, and then RTL level simulation is performed in order to validate the generated HDL architecture.

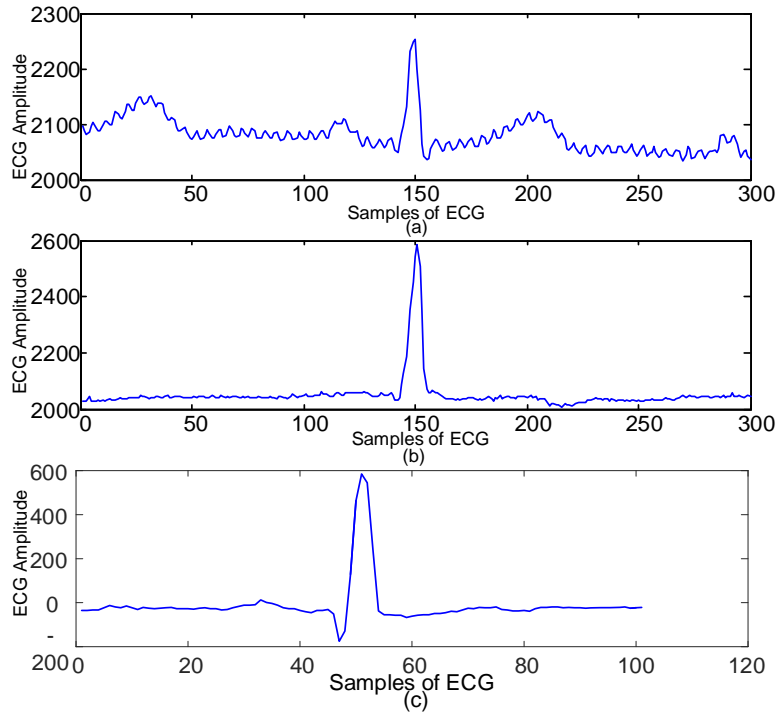

Fig 8. Samples from used ECG database. (a) Sample data collected from VS100 ECG sensor; (b) Sample collected from Shimmer ECG sensor. (c) Sample collected from MIT-BIH database.

*2) C/RTL Co-simulation*

The same C++ testbench used in the C/C++ simulation was used for the C/RTL co-simulation; however, instead of using the C++ function, the synthesized RTL architectures are used to perform the calculation. The simulator used in the C/RTL co-simulation was XSIM where VHDL was selected as the generated HDL. The clock period for the simulation was set to 10 ns. The achieved results from the C/RTL co-simulation are the same as the C/C++ simulation. Table II summaries the processing time and hardware utilization estimate of each block.

TABLE II
PROCESSING TIME AND HARDWARE UTILISATION ESTIMATE OF EACH BLOCK (C/RTL CO-SIMULATION)

| Block | Processing time (ms) | LUT (%) | FF (%) | DSP48E (%) | BRAM_18K (%) |
|---|---|---|---|---|---|
| Cipher | 0.022 | 5 | 0.8 | 0 | 3 |
| Decipher | 0.050 | 6 | 0.8 | 0 | 3 |
| ECG Identification | 0.404 | 63 | 9 | 5 | 5 |

The processing time for the Cipher and Decipher blocks is obtained for processing 128-bit input data. However, the input data size of the ECG identification block is 300×32 bits. Based on the estimated utilization of each block, it is possible to implement all three blocks in the same Zynq7 SoC, to be a complete security solution for connected health systems.

*B. Hardware Implementation*

The proposed system has been successfully implemented on the PL of the Xilinx ZC702 evaluation board. In addition, the corresponding software (i.e. Drivers and control codes) is also implemented using the Xilinx Software Development Kit running on the ARM cortex-A9 core of Zynq7 SoC.

*1) Programmable Logic Utilization*

The proposed architecture consumes about 30% of the available LUTs and 11% flip-flops. Most of LUTs, flip-flops and DSP48E are used for creating the instances of the proposed architecture, for example, AXI interfaces, multipliers, etc. Other LUTs and flip-flops are used as multiplexers or registers and for creating a memory to store the resulting matrix. It is worth noting that the target Zynq SoC has the smallest chip capacity in its family, which means that the proposed architecture has very efficient size, and can be easily deployed on a low-cost FPGA or integrated with other biometric identification systems on a large chip. Table III shows the PL hardware usage of the proposed architecture. Table IV summarizes the hardware resources usage of individual blocks.

*2) Power Consumption*

The on-chip power consumption consists mainly of two parts, which are static and dynamic power consumption. The static power is consumed due to transistor leakage. The dynamic power is consumed by fluctuating power as the design runs, i.e. Zynq7 Processing System (PS7), clock, power, logic power, signal power, BRAMs power, etc., which are directly affected by the chip clock frequency and the usage of chip area. The details of estimated power consumption of the implementation are summarised in Table V. The PS7 consumes much more power than the PL; this is due to the fact that the ARM dual core Cortex-A9 based processing system has much higher running frequency than the PL and it runs drivers and control programmes. Compare to the PS7, the custom logic blocks consumes only a small portion of the total on-chip power consumption.

TABLE III
OVERALL HARDWARE RESOURCES USAGE

| Name | Usage | Total Available Resources on Chip | Utilisation (%) |
|---|---|---|---|
| LUT | 16,133 | 53,200 | 30.3 |
| FF | 11,797 | 106,400 | 11.1 |
| BRAM_18K | 17 | 140 | 12.1 |
| DSP48E | 12 | 220 | 5.5 |

TABLE IV
HARDWARE RESOURCES USAGE FOR INDIVIDUAL CUSTOM LOGIC

| Name | LUT | FF | BRAM_18K | DSP |
|---|---|---|---|---|
| Cipher | 1,126 | 841 | 4.5 | 0 |
| Decipher | 1,212 | 829 | 5 | 0 |
| ECG Identification | 13,189 | 9,368 | 7.5 | 12 |

TABLE V
ESTIMATION OF POWER CONSUMPTION

|  | Utilization Details | Power (W) | Utilization (%) |
|---|---|---|---|
| Dynamic Power Consumption | Clock | 0.024 | 1 |
|  | Signals | 0.028 | 2 |
|  | Logic | 0.022 | 1 |
|  | DSP | 0.004 | < 1 |
|  | BRAM | 0.028 | 2 |
|  | PS7 | 1.564 | 93 |
| Static Power Consumption | Device Static | 0.158 | 9 |

*3) Timing Analysis*

The ARM processor runs at 650 MHz and the PL clocked at 50 MHz. The processing time of the proposed system is measured by counting the number of ARM processor's clock cycles spent for obtaining the calculated results of one ECG signal (i.e. 300×32 bits) from individual custom logic. Table VI shows the comparison between the software and hardware implementations of each block in terms of the processing time. As it can be seen in Table VI, the overall processing time using the hardware implementation has improved by a factor of 44 compared to the software implementation. In addition, the proposed implementation has also outperformed the existing ECG identification implementations [31] on a HS/SW hybrid platform by a factor of 305. As a result of this improvement, the processing speed meets the minimum time constraint of a real-time data processing system.

TABLE VI
PROCESSING TIME OF EACH INDIVIDUAL BLOCK

|  | Cipher | Decipher | ECG Identification | Total |
|---|---|---|---|---|
| Hardware Implementation (ms) | 3.21 | 7.41 | 0.09 | 10.71 |
| Software Implementation (ms) | 108.31 | 361.09 | 0.32 | 469.72 |

*C. Comparison with Existing Work*

This section compares the data processing throughput of the proposed cipher and decipher with the work presented in [31] and [34], as well as other existing AES-128 encryption FPGA implementation. The comparison is made based on the following three metrics: processing speed, area analysis, power consumption and overall evaluation.

*1) Comparison of the AES processing speed*

The throughput of the AES cipher and decipher is calculated using the running frequency and the execution time. The execution time is measured by the clock counter in the Zynq7 PS. The running frequency $f$ and the throughput $T$ are then calculated according to equations 3 and 4 respectively.

$$f = \frac{1}{t} \tag{3}$$

$$T = B \times f \tag{4}$$

where $t$ is the execution time and $B$ is block size which is equal to 128 bits for our case. Table VII presents the results of the proposed and exiting work.

As it can be seen from Table VII, the throughputs of the proposed AES cipher and decipher implementations have outperformed the existing work in [31] and [34] by a factor of 1.6 and 3.7 respectively. This is due to the fact that the proposed implementation has used appropriate directives in Vivado HLS in order to reduce the latency of the proposed architectures.

TABLE VII
COMPARISON OF AES PROCESSING SPEED

|  | Cipher | | | Decipher | | |
| --- | --- | --- | --- | --- | --- | --- |
|  | [31] | [34] | Proposed work | [31] | [34] | Proposed work |
| Execution time (µs) | 41.0 | 40.8 | 25.1 | 213 | 211 | 57.9 |
| Execution frequency (kHz) | 24.4 | 24.5 | 39.9 | 4.7 | 4.7 | 17.3 |
| Throughput (Gbit/s) | 3.1 | 3.1 | 5.1 | 0.6 | 0.6 | 2.2 |

*2) Comparison of Hardware Resources Usage for AES implementation*

The occupied area after placement and routing in terms of slices, LUTs and BRAMs is shown in Table VIII. The proposed work has significantly less hardware resource requirement to implement AES cipher and decipher block.

TABLE VIII

COMPARISON OF HARDWARE RESOURCES FOR AES IMPLEMENTATION

|  | Cipher | | | Decipher | | |
|---|---|---|---|---|---|---|
|  | [31] | [34] | Proposed work | [31] | [34] | Proposed work |
| Slice | 3443 | 3430 | 372 | 5723 | 5536 | 431 |
| LUTs | 5513 | 8589 | 1126 | 5514 | 8262 | 1212 |
| BRAM | 2 | 2 | 4.5 | 2 | 1 | 5 |

*3) Comparison of Power Consumption for AES implementation*

Table IX shows the comparison of power consumption for AES implementation. According to the results shown in Table IX, the proposed implementation consumes less power than the existing work. This improvement has not only benefited from the new technologies introduced in the latest SoC, but also is optimized using Vivado HLS.

TABLE IX
COMPARISON OF POWER CONSUMPTION FOR AES IMPLEMENTATION

|  | Cipher | | | Decipher | | |
|---|---|---|---|---|---|---|
|  | [31] | [34] | Proposed work | [31] | [34] | Proposed work |
| Clock (mW) | 45 | 45 | 2 | 95 | 92 | 2 |
| Logic (mW) | 10 | 10 | 2 | 48 | 35 | 3 |
| Signals (mW) | 25 | 25 | 2 | 252 | 192 | 2 |
| Total (mW) | 132 | 132 | 6 | 447 | 371 | 7 |

*4) Comparison of Overall AES-128 encryption implementation*

Table X shows the comparison of overall AES-128 encryption implementations of the proposed work with other existing works. As it can be seen from Table X, the proposed work outperforms other existing work in terms of Throughput/Slice, which achieves 13.73 Mbps/Slice. Although the proposed work does not have the highest throughput in the comparison, but it uses the lowest number of slices of the available chip area. In addition, the proposed implementation has significantly lower the clock speed compared to other work, which means that it would have better power efficiency.

TABLE X
COMPARISON OF AES-128 ENCRYPTION FPGA IMPLEMENTATION

| Work | Device | Clock (MHz) | Throughput (Mbps) | Area (Slices) | Throughput /Area (Mbps/Slice) |
|---|---|---|---|---|---|
| Proposed work | Zynq XC7Z020 | 50 | 5107 | 372 | 13.73 |
| [14] | Virtex-V XC5LVX85 | 576.07 | 73737 | 22994 | 3.21 |
| [35] | Spartan-III XC3S4000 | 206.28 | 2640 | 405 | 6.520 |
| [36] | Virtex-IV XC4VLX100 | 645.70 | 82650 | 12256 | 6.744 |
| [37] | Virtex-IV XC4VLX25 | 166.7 | 2134 | 626 | 4.41 |
| [38] | APEX20KC | N/A | 1188 | 895 | 1.33 |

*5) Comparison of Overall ECG Identification implementation*

Table XI shows the comparison of the proposed FPGA-based ECG identification implementation with other existing work. The proposed implementation processes a sample in 0.09 ms which outperforms other existing work. Although the proposed work uses more LUTs for the implementation as maximized pipeline and parallelism were set to achieve a high processing speed that meet the real-time requirements.

TABLE XI
COMPARISON OF FPGA BASED ECG IDENTIFICATION IMPLEMENTATION

| Work | Device | Clock (MHz) | Processing Speed (ms) | LUTs |
|---|---|---|---|---|
| Proposed work | Zynq XC7Z020 | 50 | 0.09 | 13,189 |
| [39] | Virtex-II | 20 | 1.87 | 2170 |
| [40] | Spartan 3 | 36 | N/A | 9,136 |

VI. CONCLUSION

In this paper a set of security solutions have been presented that ensure patients' data and medical records within a connected health system can be securely transmitted and saved for further analysis and diagnosis. The proposed ECG identification system consists of a SoC implementation of the AES and ECG identification algorithms, which can be used to collect data from ECG sensors, process it and meet the real-time response constraint. The proposed Zynq SoC implementation provides not only high data processing performance but they also provide the user with higher flexibility in terms of various interfaces, hard processors and PL. Results presented have shown that the proposed implementation only needs 30% of hardware resources and 107 mW to process an ECG sample in 10.71 ms, which outperforms the existing FPGA-based system in different key performance metrics.